\documentclass[aps,twocolumn,floatfix,groupedaddress,superscriptaddress,longbibliography,footinbib,notitlepage,showpacs]{revtex4-1}
\usepackage{graphicx,graphics,times,bm,bbm,bbold,amssymb,soul}
\usepackage[T1]{fontenc}
\usepackage{amsthm,amsmath,amsfonts,dsfont,color,xcolor,dcolumn}
\usepackage[bookmarksnumbered,colorlinks,plainpages]
{hyperref}
\hypersetup{
     colorlinks = true,
     citecolor  = blue,
     urlcolor   = blue}
     
\newcolumntype{P}[1]{>{\centering\arraybackslash}p{#1}}

\usepackage{tabularx,ragged2e}

\newtheorem{manualtheoreminner}{Theorem}
\newenvironment{manualtheorem}[1]{%
  \renewcommand\themanualtheoreminner{#1}%
  \manualtheoreminner
}{\endmanualtheoreminner}

\newtheorem{manualpropositioninner}{Proposition}
\newenvironment{manualproposition}[1]{%
  \renewcommand\themanualpropositioninner{#1}%
  \manualpropositioninner
}{\endmanualpropositioninner}

\newtheorem{manualconjectureinner}{Conjecture}
\newenvironment{manualconjecture}[1]{%
  \renewcommand\themanualconjectureinner{#1}%
  \manualconjectureinner
}{\endmanualconjectureinner}

\usepackage{soul}

\begin{document}

\title{An Algebraic-Geometric Characterization of Tripartite Entanglement}

\author{Masoud Gharahi} 
\email[]{masoud.gharahi@gmail.com}
\email[]{masoud.gharahighahi@unicam.it}
\affiliation{School of Science and Technology, University of Camerino, 62032 Camerino, Italy}
\affiliation{INFN Sezione di Perugia, 06123 Perugia, Italy}
\author{Stefano Mancini}
\email[]{stefano.mancini@unicam.it}
\affiliation{School of Science and Technology, University of Camerino, 62032 Camerino, Italy}
\affiliation{INFN Sezione di Perugia, 06123 Perugia, Italy}


\begin{abstract}
To characterize entanglement of tripartite $\mathbbm{C}^d\otimes\mathbbm{C}^d\otimes\mathbbm{C}^d$ systems, we employ algebraic-geometric tools that are invariants under Stochastic Local Operation and Classical Communication (SLOCC), namely $k$-secant varieties and one-multilinear ranks. Indeed, by means of them, we present a classification of tripartite pure states in terms of a finite number of families and subfamilies. At the core of it stands out a fine-structure grouping of three-qutrit entanglement.
\end{abstract}


\maketitle

\section{Introduction}

Entanglement, besides its foundational relevance, has been recently recognized as a fundamental resource for quantum information processing \cite{HHHH09}. Therefore its characterization became of uppermost importance. For bipartite systems this was done by developing entanglement monotones \cite{HHHH09}. However extension of these to multipartite systems soon appeared quite challenging. 
That is why a classification of entangled states in multipartite systems was pursued on the basis of one out of the many properties satisfied by entanglement monotones, namely the invariance under local operation and classical communication. 
Actually, this property is reinforced by requiring stochasticity of local operation and classical communication (SLOCC).  
Such invariance property is relevant to single out states that perform (probabilistically) equally well quantum information tasks. Along this way there have been several attempts to classify multiqubit pure states \cite{DVC00, VDDV02, BDDMR10, RM11, Aulbach12, LL12, GW13, GA16, GM18}. In effect, many approaches were put forward on the benchmarks of three or four qubits, but were not extensible to a larger number of qubits. Recently we proposed an algebraic-geometric approach that works for an arbitrary number of qubits, although with a hard computational complexity \cite{GMO20}.

Going beyond qubit, more information can be encoded in qudits and more robustness against noise can be achieved \cite{CMM12}. Also, in quantum cryptography, entangled qudits guarantee more security against coherent attacks than using entangled qubits \cite{CBKG02}. These facts motivate the classification of entangled states for higher than two-dimensional systems. Ref \cite{YWWF08}, has investigated the SLOCC classification of two- and three-qutrit entanglement based on the inductive method. However, this method suffer from a flaw already at qubits level \cite{GM18}, which propagates to higher dimensional systems \cite{example}. In Ref. \cite{BLTV04}, the invariants of three-qutrit entanglement has been studied, while Ref. \cite{HJ16} used singularity theory to study the entanglement of pure three-qutrit states. More specifically, Refs. \cite{Nurmiev1,Nurmiev2} provided an implicit description of all three  fundamental invariants of ${\rm{SL}}(3,\mathbbm{C})^{\times{3}}$, and classified the normal forms in five families, which can also be derived as a special case of entanglement classification of three-fermions with nine single-particle states \cite{SL14}.

Other promising approaches that might be employed to characterize entanglement of qudits states are those based on symplectic and algebraic geometry \cite{BH01, Miyake03,BGH07, SHK11, HLT12, ST13, WDGC13, SOK14, HLT14-17, SMKKKO18}. In the symplectic-geometric approach, momentum map (which relates a state to its one-particle reduced density matrices) and its norm (which is related to the linear entropy) are the key tools \cite{SMKKKO18}, while in the algebraic-geometric approach, the Segre, Veronese, and Pl\"{u}cker embedding maps together with their secant varieties are the key tools (to be used with distinguishable particles, bosons and fermions, respectively) \cite{HLT14-17,GMO20}. Regarding the momentum map for the SLOCC classification, the main obstacle is that SLOCC do not preserve the symplectic form \cite{SMKKKO18}. Therefore, we pursue the extendibility of the algebraic-geometric approach of Ref. \cite{GMO20} to multiqudit states using as a benchmark tripartite systems and achieving in particular a full classification of three-qutrit states. Moving on from the space $\mathbbm{C}^d\otimes\mathbbm{C}^d\otimes\mathbbm{C}^d$, we shall show that the class of fully separable states corresponds to a Segre variety, and that the $k$-secant varieties constructed from it are SLOCC invariants, which gather infinite (actually uncountable) number of SLOCC classes to a finite number of entanglement families. Additionally, we employ one-multilinear ranks (hereafter one-multiranks) to divide each SLOCC family into a finite number of SLOCC subfamilies. On the one hand, a $k$-secant of the Segre variety joins its $k$ points which corresponds to an entangled state being a superposition of $k$ fully separable states. On the other hand, one-multiranks are triples of integers which are just ranks of different flattenings of a given tripartite state as an order-3 tensor in $\mathcal{H}={\mathbbm{C}^d}^{\otimes{3}}$.

The paper is organized as follows. Sec. \ref{Sec.ii}, is devoted to present tools from algebraic geometry that are SLOCC invariants.  Also the entanglement classification algorithm based on them will be provided. Then, in Sec. \ref{Sec.iii}, we study in details the three-qutrit entanglement achieving a fine-structure classification as a relevant example. Several of issues of this case will be generalized to qudit systems in Sec. \ref{Sec.iv}. Finally, we draw our conclusions and discuss future perspectives in Sec. \ref{Sec.v}.

\section{The classification method}\label{Sec.ii}

At the core of Ref. \cite{GMO20} was the identification of determinantal and secant Segre varieties as those that classify multiqubit entanglement. Here, we shall extend this approach to classify three-qudit pure states
\begin{equation}\label{3quditstate}
|\psi\rangle = \sum_{i\in\{0,\ldots,d-1\}^{3}}\mathfrak{c}_{i}|i\rangle\,.
\end{equation}
To this end, we shall be examining maps $\mathcal{M}$ that are produced by tensor flattening \cite{Landsberg} from the quantum states in Eq. (\ref{3quditstate}). Consider the tensor Hilbert space 
$\mathcal{H}=\mathcal{H}_1\otimes\mathcal{H}_2\otimes\mathcal{H}_3$, where $\mathcal{H}_i\simeq \mathbbm{C}^d$. We shall define $\ell$-partitions as ordered $\ell$-tuples $I=\left(i_1,\ldots,i_{\ell}\right)$, where $1\leq\ell\leq{2}$, and $1\leq i_1<\cdots<i_{\ell}\leq{3}$. Given an $\ell$-partition $I$, we define the complementary partition $\bar{I}$ as the $(3-\ell)$-partition such that $I\cup\bar{I}=\{1,2,3\}$. Therefore, $\mathcal{H}\simeq\mathcal{H}_I\otimes\mathcal{H}_{\bar{I}}$, where $\mathcal{H}_I={\mathbbm{C}^{d}}^{\otimes\ell}$ and $\mathcal{H}_{\bar{I}}$ is the complementary Hilbert space. For any state $\psi$ with vector representation $|\psi\rangle\in\mathcal{H}$, the $\ell$-partition $I$ leads to a linear operator $\mathcal{M}_{I}[\psi]$ which maps 
the dual $\mathcal{H}_I^*$ of ${\cal H}_I$ to ${\cal H}_{\bar{I}}$,
\begin{equation}\label{flattening}
\mathcal{M}_I[\psi]:\mathcal{H}_I^*\to\mathcal{H}_{\bar{I}}\,,
\end{equation}
as, in Dirac notation,
\begin{equation}
\mathcal{M}_{I}[\psi]=\left(\langle{e_0}|\psi\rangle,\ldots,\langle{e_{d^{\ell}-1}}|\psi\rangle\right)^{\rm{T}}\,,
\end{equation}
where $|e_j\rangle=|j\rangle$ is the computational basis of $\mathcal{H}_{I}$ and $\rm{T}$ denotes the matrix transposition. Clearly, we shall consider all ordered $\ell$-tuples $I$ to avoid overlapping of entanglement families \cite{GM18}. Given a state $\psi$ and a number $1\leq\ell\leq{2}$, we call the sequence of ranks $r_{I}[\psi]={\rm{rank}}\left(\mathcal{M}_I[\psi]\right)$ for all $\ell$-partitions $I$, the $\ell$-multilinear rank (hereafter $\ell$-multirank) of the state $\psi$. Although there are six partitioning, with three complementary pairs $(1) \leftrightarrow (23)$, $(2) \leftrightarrow (13$), $(3) \leftrightarrow (12)$, it is enough to check $\ell$-multiranks for partition $I$ with $\ell=1$. Note that for the complementary partition $\bar{I}$ the matrices $\mathcal{M}_{\bar{I}}[\psi]$ are just the transpose of $\mathcal{M}_{I}[\psi]$ and transposition does not change the rank of the matrix.

An important observation is that, for any partition $I$, the rank of the linear map $\mathcal{M}_I[\psi]$ is the same as the rank of the reduced density matrix obtained after tracing over the parties identified by the complementary partition $\bar{I}$, i.e.,
\begin{equation}
\rho_I={\rm{Tr}}_{\bar{I}}\left[|\psi\rangle\langle\psi|\right]=\mathcal{M}_I[\psi]\mathcal{M}_{\bar{I}}^{\dagger}[\psi]\,.
\end{equation}

Another important observation is that SLOCC equivalent states, i.e.,
\begin{equation}
|\tilde{\psi}\rangle = \left(\otimes_{i=1}^{3} A_{i}\right) |\psi\rangle\,,
\end{equation}
where $|\psi\rangle\in\mathcal{H}$ and $A_{i}\in{\rm{SL}}(d,\mathbbm{C})$, yield
\begin{equation}
\mathcal{M}_{I}[\tilde{\psi}]=\left(\otimes_{i\in I}A_{i}\right)\mathcal{M}_{I}[\psi]\left(\otimes_{i\in\bar{I}}A_{i}\right)^{\rm{T}}\,.
\end{equation}
Therefore, $\ell$-multirank is an invariant under SLOCC.

Since $\ell$-multiranks only depend on the quantum state, and not on the representation, and, furthermore, because statements about rank can be rephrased as statements about minors which are determinants, it follows that a given $\ell$-multirank configuration determines a determinantal variety in the projective Hilbert space $\mathbbm{P}\mathcal{H}$. Actually, the determinantal variety is a subset of all matrices with rank $r$ or less in $\mathbbm{P}\mathcal{H}$, that is just the common zero locus of the $(r+1)\times(r+1)$ minors. Tripartite pure states which have $\ell$-multiranks bounded by a given integer sequence make a subvariety of $\mathbbm{P}\mathcal{H}$. In particular, the Segre variety is an example of a determinantal variety; it is the zero locus of the $2\times2$ minors of the coefficient matrices in Eq. \eqref{3quditstate}, i.e., common zero locus of the quadratic polynomials $\mathcal{M}_{ij}\mathcal{M}_{kl}-\mathcal{M}_{il}\mathcal{M}_{kj}$. Therefore, the projective variety of fully separable three-qudit states has the structure of a Segre variety which is embedded in the ambient space as follows:
\begin{equation}
\Sigma^{3}_{\textbf{d-1}}:~\mathbbm{P}^{d-1}\times\mathbbm{P}^{d-1}\times\mathbbm{P}^{d-1}\hookrightarrow\mathbbm{P}^{d^{3}-1}\,.
\end{equation}
Here, $\textbf{d-1}=(d-1,d-1,d-1)$ and $\times$ is the Cartesian product of sets. One can readily check that $\Sigma^{3}_{\textbf{d-1}}$ is indeed the projective variety of fully separable states. Actually, if all partial traces are pure states, the corresponding ranks are all one. So we have that for all $\ell$-partitions the rank of 
$\mathcal{M}_I[\psi]$ is always one. Conversely, if all ranks are one, the state is fully separable.

The join of two projective varieties $\mathcal{X}$ and $\mathcal{Y}$, where both are subvarieties of a projective variety, is given by the algebraic closure, for the Zariski topology, of the lines from one to the other,
\begin{equation}
\mathfrak{J}(\mathcal{X},\mathcal{Y})=\overline{\bigcup_{x\in\mathcal{X},y\in\mathcal{Y},x\neq{y}}\mathbbm{P}^1_{xy}}\,,
\end{equation}
where $\mathbbm{P}^1_{xy}$ is the projective line that includes both $x$ and $y$. If $\mathcal{X}=\mathcal{Y}$, the joining is called the secant variety of $\mathcal{X}$, i.e., $\sigma(\mathcal{X})=\mathfrak{J}(\mathcal{X},\mathcal{X})$. Therefore, the secant varieties are given by the algebraic closure of the joining of the Segre variety and the immediately previous secant variety, i.e.,
\begin{equation}
\sigma_{k}(\Sigma)=\mathfrak{J}\left(\sigma_{k-1}(\Sigma),\Sigma\right)\,.
\end{equation}
with the formal definition that $\sigma_1(\Sigma)=\Sigma$. This means that a generic point of the $k$-secant variety is a combination of $k$ distinct points of the Segre variety (the superposition of $k$ fully separable states), whence we say that the generic tensor rank is $k$. A crucial element of the definitions is that the secants are closed. This means that in each $k$-secant family there will be elements whose tensor rank will not be $k$ (it can be greater than $k$) but the tensor border rank (border rank, for short) is $k$. Thus we will make the distinction between the proper secant and the tangent.

Note that, on the one hand, the rank of a tensor $\psi$ is defined as the minimum number of simple tensors (fully separable states) that sum to $\psi$ and it extends the notion of the rank of a matrix in algebra \cite{Bourbaki}, so it can be seen as a generalization of Schmidt rank. On the other hand, the border rank of a tensor $\psi$ is defined as the smallest $r$ such that $\psi$ is a limit of tensors of rank $r$, or equivalently the smallest $r$ such that $\psi$ lies in the Zariski closure of the set of tensors of rank $r$ \cite{Landsberg}, so it can be seen as a counterpart of the generalized Schmidt rank. Interested reader can learn more about the tensor rank in the context of quantum entanglement in Ref. \cite{BFZ20}.

Suppose now $\mathcal{Y}\subset\mathcal{X}$ and let tangent star 
$\mathcal{T}^{\star}_{\mathcal{X},\mathcal{Y},y_{0}}$ denotes the union of $\mathbbm{P}_{\star}^{1}=\lim_{x,y\to{y_{0}}}\mathbbm{P}^{1}_{xy}$ with $y_{0}\in\mathcal{Y}$. The variety of relative tangent star is defined as follows:
\begin{equation}
\mathcal{T}(\mathcal{X},\mathcal{Y})=\bigcup_{y\in\mathcal{Y}}\mathcal{T}^{\star}_{\mathcal{X},\mathcal{Y},y}\,.
\end{equation}
We denote the tangential variety as $\tau(\mathcal{X})=\mathcal{T}(\mathcal{X},\mathcal{X})$. Obviously $\tau_{k}(\Sigma)\subset\sigma_{k}(\Sigma)$ and $\mathcal{T}(\tau_{k-1}(\Sigma),\Sigma)\subset\tau_{k}(\Sigma)$, the last inclusion is even an equality.

It worth noting that in addition to the standard flattenings, as the standard tensor contraction shown in Eq. (\ref{flattening}), for tripartite systems $\mathbbm{C}^{2m+1}\otimes\mathbbm{C}^{2m+1}\otimes\mathbbm{C}^{2m+1}$, we have another flattening map as follows:
\begin{equation}\label{flattening2}
\Lambda^{m}\mathcal{H}_1\otimes\mathcal{H}^*_2\to\Lambda^{m+1}\mathcal{H}_1\otimes\mathcal{H}_3\,,
\end{equation}
where $\Lambda^{m}$ denotes the $m^{\text{th}}$ exterior power. Hence, the size $(k+1)\binom{2m}{m}$ minors of Eq. \eqref{flattening2} provides equations for $k$-secant variety up to $k=(2m+1)^2/(m+1)$ (see Refs. \cite{Landsberg, LO13}).

If the points of a given variety $\mathcal{X}$ remains invariant under the action of a group, then so is any of its auxiliary variety which is built from points of $\mathcal{X}$. It means that the $k$-secant variety of Segre variety is invariant under the action of the projective linear group and therefore is an SLOCC invariant. Therefore, SLOCC classes gather naturally into entanglement families. For this reason, the dimension of the higher $k$-secant variety, which fills the projective Hilbert space of three qudits, can indicate the number of entanglement families.
The higher $k$-secant variety fills the ambient space $\mathbbm{P}(\mathbbm{C}^{d}\otimes\mathbbm{C}^{d}\otimes\mathbbm{C}^{d})$ when
\begin{equation}\label{TripartiteGenericRank}
k=\left\lceil\frac{d^3}{3d-2}\right\rceil\,,
\end{equation}
except for $d=3$ where the generic rank is five \cite{Lickteig,Strassen}. This $k$ indicates the number of entanglement families which remains finite with the dimension of parties.

Since $\sigma_{k-1}\subset\sigma_k$ we need to distinguish the elements of each $k$-secant family by defining the proper secant. The proper $k$-secant [the states that belongs to $k$-secant but not to $(k-1)$-secant], i.e., the set $\sigma_{k}(\Sigma^{3}_{\textbf{d-1}})\setminus\sigma_{k-1}(\Sigma^{3}_{\textbf{d-1}})$, is the union of the $k$-secant hyperplanes $\mathcal{S}_{k}\subset\sigma_{k}(\Sigma^{3}_{\textbf{d-1}})$ represented by
\begin{equation}\label{S-plane}
\mathcal{S}_{k}=\sum_{i=1}^{k}\lambda_{i}p_{i}\,,
\end{equation}
with $\{\lambda_{i}\}_{i=1}^{k}\neq{0}$ and each $p_{i}$ is a distinct point in Segre variety.

Therefore, similar to the spirit of Ref. \cite{GMO20}, we use \mbox{$k$-secant} varieties and one-multiranks as the SLOCC invariants to bunch entanglement orbits (classes) of tripartite 
${\mathbbm{C}^d}^{\otimes{3}}$ systems into a finite number of families and subfamilies. Hence, the classification algorithm can be summarized as:
\begin{itemize}
\item[{\bf (i)}] find families by identifying $k$-secant varieties $\Sigma^{3}_{\textbf{d-1}}, \sigma_{2}(\Sigma^{3}_{\textbf{d-1}}), \ldots, \sigma_{k}(\Sigma^{3}_{\textbf{d-1}})$;
\item[{\bf (ii)}] split families to secants and tangents by identifying $\tau_{2}(\Sigma^{3}_{\textbf{d-1}}), \ldots, \tau_{k}(\Sigma^{3}_{\textbf{d-1}})$;
\item[{\bf (iii)}] find subfamilies by identifying one-multiranks.
\end{itemize}

\section{Fine-structure classification of three-qutrit entanglement}\label{Sec.iii}

For the Segre surface $\Sigma^{3}_{\textbf{2}}\subset\mathbbm{P}^{26}$, we shall use homogeneous coordinates associated with the induced basis $\left\{|000\rangle,|001\rangle,\ldots,|222\rangle\right\}$. That is to say, a point $p\in\mathbbm{P}^{26}$ is written in homogeneous coordinates $\left[\mathfrak{c}_0:\mathfrak{c}_1:\cdots:\mathfrak{c}_{26}\right]$ whenever $p$ is the projective class of the three-qutrit state of Eq. (\ref{3quditstate}). Then, the Segre surface $\Sigma^{3}_{\textbf{2}}$ is the projective variety with points given by affine coordinates $[1:a:b:c:ac:bc:d:ad:bd:e:ae:be:ce:ace:bce:de:ade:
bde:f:af:bf:cf:acf:bcf:df:adf:bdf]$, where $a$, $b$, $c$, $d$, $e$, and $f$ are complex parameters. This expression must be properly understood, in that the limits of $a$ and/or $b$ and/or $c$ and/or $d$ and/or $e$ and/or $f$ going to infinity, must be included. For instance, also points of the form $[0:1:0:0:c:0:0:d:0:0:e:0:0:ce:0:0:de:0:0:f:0:0:cf:0:0:df:0]$, which corresponds to $a\rightarrow\infty$, are part of $\Sigma^{3}_{\textbf{2}}$.

Thanks to Ref. \cite{CK11}, all one-multiranks can be found for states of any number of qudits. For three-qutrit states we have
\begin{equation}
r_i\leq\prod_{j\neq{i}}r_j\quad\forall~i,j\in\{1,2,3\}\,,
\end{equation}
where $0\leq{r_i}\leq{3}$ stands for the rank of the corresponding flattening. Therefore, all the 
one-multiranks of three-qutrit states are: (111) which indicates a fully separable states; (122) and (133) and their permutations, which indicate biseparable states; (222), all permutations of (223), all permutations of (233), and (333), which indicate genuinely entangled states.

Standard flattenings are not enough to construct higher secant families in $\mathbbm{P}^{26}$. So based on Eq. \eqref{flattening2} we have the following flattening:
\begin{equation}\label{flat2}
\mathcal{F}:~\mathcal{H}_1\otimes\mathcal{H}^*_2\to\Lambda^2\mathcal{H}_1\otimes\mathcal{H}_3\,,
\end{equation}
that can be considered as the composition of
$$
\mathcal{H}_1\otimes\mathcal{H}^*_2\xrightarrow[]{{\rm{Id}}_{\mathcal{H}_1}\otimes\mathcal{M}_2}\mathcal{H}_1\otimes\mathcal{H}_1\otimes\mathcal{H}_3\,,
$$
and
$$
\mathcal{H}_1\otimes\mathcal{H}_1\otimes\mathcal{H}_3\xrightarrow[]{P_\wedge\otimes{\rm{Id}}_{\mathcal{H}_3}}\Lambda^2\mathcal{H}_1\otimes\mathcal{H}_3\,,
$$
where $\mathcal{M}_2:\mathcal{H}^*_{2}\to\mathcal{H}_1\otimes\mathcal{H}_3$ is the standard flattening and $P_\wedge:\mathcal{H}_1\otimes\mathcal{H}_1\to\Lambda^2\mathcal{H}_1$ is the projection onto the skew-symmetric component \cite{Ottaviani09}. Based on the map in Eq. \eqref{flat2}, we have the following $9\times{9}$ matrix (known as Ottaviani-Strassen matrix) for the general three-qutrit state of Eq. (\ref{3quditstate}),
\begin{align}\nonumber
&\mathcal{F}= \\ \label{strassen}
&\left(
\begin{array}{ccccccccc}
 0 & 0 & 0 & \mathfrak{c}_{0} & \mathfrak{c}_{1} & \mathfrak{c}_{2} & -\mathfrak{c}_{9} & -\mathfrak{c}_{10} & -\mathfrak{c}_{11} \\
 0 & 0 & 0 & \mathfrak{c}_{3} & \mathfrak{c}_{4} & \mathfrak{c}_{5} & -\mathfrak{c}_{12} & -\mathfrak{c}_{13} & -\mathfrak{c}_{14} \\
 0 & 0 & 0 & \mathfrak{c}_{6} & \mathfrak{c}_{7} & \mathfrak{c}_{8} & -\mathfrak{c}_{15} & -\mathfrak{c}_{16} & -\mathfrak{c}_{17} \\
-\mathfrak{c}_{0} & -\mathfrak{c}_{1} & -\mathfrak{c}_{2}  & 0 & 0 & 0 & \mathfrak{c}_{18} & \mathfrak{c}_{19} & \mathfrak{c}_{20} \\
-\mathfrak{c}_{3} & -\mathfrak{c}_{4} & -\mathfrak{c}_{5} & 0 & 0 & 0 & \mathfrak{c}_{21} & \mathfrak{c}_{22} & \mathfrak{c}_{23} \\
-\mathfrak{c}_{6} & -\mathfrak{c}_{7} & -\mathfrak{c}_{8} & 0 & 0 & 0 & \mathfrak{c}_{24} & \mathfrak{c}_{25} & \mathfrak{c}_{26} \\
\mathfrak{c}_{9} & \mathfrak{c}_{10} & \mathfrak{c}_{11} & -\mathfrak{c}_{18} & -\mathfrak{c}_{19} & -\mathfrak{c}_{20} & 0 & 0 & 0 \\
\mathfrak{c}_{12} & \mathfrak{c}_{13} & \mathfrak{c}_{14} & -\mathfrak{c}_{21} & -\mathfrak{c}_{22} & -\mathfrak{c}_{23} & 0 & 0 & 0 \\
\mathfrak{c}_{15} & \mathfrak{c}_{16} & \mathfrak{c}_{17}  & -\mathfrak{c}_{24} & -\mathfrak{c}_{25} & -\mathfrak{c}_{26} & 0 & 0 & 0
\end{array}
\right)\,.
\end{align}
Actually, the determinant of matrix $\mathcal{F}$, which is an ${\rm{SL}}(3,\mathbbm{C})^{\times{3}}$-invariant of degree nine, indicates the four-secant hyperplane. It means that if $\mathcal{F}$ is full rank for a given state, i.e., rank of the matrix $\mathcal{F}$ is nine, that state is in five-secant family. Indeed, $\lceil\frac{\text{rank}\mathcal{F}}{2}\rceil$ indicate the secant family of the state.

Let us now move on to the proper two-secant variety, i.e., the set $\sigma_{2}(\Sigma^{3}_{\textbf{2}})\setminus\Sigma^{3}_{\textbf{2}}$, which is the union of the secant planes $\mathcal{S}_{2}$ represented by Eq. (\ref{S-plane}). The generic element in the proper two-secant comes from joining two distinct points (superposition of two fully separable states), i.e., $\lambda_1 p_1+\lambda_2 p_2$ with $\lambda_1,\lambda_2\neq{0}$. For instance, it is easy to see that
\begin{equation}\label{GHZ3-1}
|{\rm{GHZ}}_3^{(1)}\rangle=|\alpha\alpha\alpha\rangle+|\beta\beta\beta\rangle\,,
\end{equation}
where $\alpha\neq\beta\in\{0,1,2\}$ is an element of $\sigma_{2}(\Sigma^{3}_{\textbf{2}})$ with one-multirank equal to $(222)$. 

Also, the classification of two-qutrit states (see  Appendix \ref{AppxA}) provides us the following biseparable states with one-multirank equal to $(122)$, up to a permutation, as other elements of 
$\sigma_{2}(\Sigma^{3}_{\textbf{2}})$:
\begin{equation}
|{\rm{B}}_{i}^{(1)}\rangle_{i=1}^{3}=\mathcal{P}\{|{\rm{GHZ}}_{2}^{(1)}\rangle|\text{1-qutrit}\rangle\}\,,
\end{equation}
where $\mathcal{P}\{\cdot\}$ denotes all possible permutations of subsystems and similarly to Eq. \eqref{GHZ3-1} $|{\rm{GHZ}}_{2}^{(1)}\rangle=|\alpha\alpha\rangle+|\beta\beta\rangle$. Note that this is the situation in which one or more parameters on the proper two-secant variety tend to infinity.

Now, considering the tangent to the point $p_1=[1:0:\cdots:0]$ (equivalent to all points on $\Sigma^{3}_{\textbf{2}}$ by an SLOCC), we have the affine coordinate $[1:\mu:\mu:\mu:0:0:\mu:0:0:\mu:0:0:0:0:0:0:0:0:\mu:0:0:0:0:0:0:0:0]$. Letting $\mu\to\infty$, we have the state $|00\upsilon \rangle+|0\upsilon0\rangle+|\upsilon00\rangle$ with $|\upsilon\rangle=|1\rangle+|2\rangle$ which is obviously a three-qutrit $\rm{W}$-type state. Bearing in mind this result, we can derive the following state as an element of $\tau_{2}(\Sigma^{3}_{\textbf{2}})$ with one-multirank equal to $(222)$:
\begin{equation}\label{W3}
|{\rm{W}}_3\rangle=|{\rm{D}}_{3}^{\mathcal{P}(2,1,0)}\rangle=\sum_{i}\mathcal{P}_{i}\{|\alpha\alpha\beta\rangle\}\,,
\end{equation}
where $\alpha\neq\beta\in\{0,1,2\}$ and
\begin{equation}\label{Dicke-3-qudit}
|{\rm{D}}_{3}^{\jmath}\rangle=\sqrt{\frac{\prod_{i}{j_i!}}{3!}}\sum_{\pi\in\mathfrak{S}_{3}}\pi\{|0\rangle^{\otimes{j_1}}\otimes\cdots\otimes|d-1\rangle^{\otimes{j_d}}\}\,,
\end{equation}
are the so-called $3$-qudit Dicke states (with excitations shown as $\jmath=(j_1,\ldots,j_d)$ where $j_1+\cdots+j_d=3$). Also, we can explicitly see that $|{\rm{W}}_3\rangle$ can asymptotically be obtained from $|{\rm{GHZ}}_3^{(1)}\rangle$ as follows:
\begin{equation}
|{\rm{W}}_3\rangle=\lim_{\varepsilon\to{0}}\frac{1}{\varepsilon}\left((|\alpha\rangle+\varepsilon|\beta\rangle)^{\otimes{3}}-|\alpha\alpha\alpha\rangle\right)\,.
\end{equation}

The proper three-secant, i.e., the set $\sigma_{3}(\Sigma^{3}_{\textbf{2}})\setminus\sigma_{2}(\Sigma^{3}_{\textbf{2}})$, is the union of the secant hyperplanes $\mathcal{S}_{3}$ represented by Eq. (\ref{S-plane}). So, joining three distinct points in the Segre variety (superposition of three fully separable states) that satisfies Eq. (\ref{S-plane}), gives rise to elements of proper three-secant family. For instance,
\begin{equation}\label{GHZ3-2}
|{\rm{GHZ}}_{3}^{(2)}\rangle=|000\rangle+|111\rangle+|222\rangle\,, 
\end{equation}
is an element of $\sigma_{3}(\Sigma^{3}_{\textbf{2}})$ with one-multirank equal to $(333)$. In the proper three-secant we have other elements with different one-multiranks. For instance,
\begin{equation}\label{(322)}
|{\rm{GHZ}}_3^{(1)}\rangle+\mathcal{P}\{|\alpha\beta\gamma\rangle\}\,,
\end{equation}
and
\begin{equation}\label{(332)}
|{\rm{GHZ}}_3^{(1)}\rangle+\mathcal{P}\{|\alpha\gamma\gamma\rangle\}\,,
\end{equation}
where $\alpha\neq\beta\neq\gamma\in\{0,1,2\}$ are all elements of $\sigma_{3}(\Sigma^{3}_{\textbf{2}})$ with one-multirank equal to $(223)$ and $(233)$, up to a permutation, respectively. The states in Eqs. \eqref{(322)} and \eqref{(332)} are the joining of a $|{\rm{GHZ}}_3^{(1)}\rangle$ state and a distinct point in the Segre variety. One can write these elements of proper three-secant in terms of joining biseparable states $|{\rm{B}}^{(1)}\rangle$ and a distinct point of Segre variety as $|\alpha\rangle(|\alpha\alpha\rangle+|\beta\gamma\rangle)+|\beta\beta\beta\rangle$ and $|\alpha\rangle(|\alpha\alpha\rangle+|\gamma\gamma\rangle)+|\beta\beta\beta\rangle$, respectively.

\begin{table*}[t]
\centering
\caption{Fine-structure classification of three-qutrit entanglement. Each column corresponds to a family ($\tau_k$ is the closure of $\sigma_k$ family that is split based on tensor rank). Within a column, each row corresponds to a subfamily. A subscript $k$ is used to indicate members appearing in different $k$-secant families while having the same one-multirank. A prime symbol is used for states in the $k$-tangent variety that appear (with the same one-multirank) in the boundary of the $k$-secant variety.}
\begin{tabularx}{\linewidth}{XXXXXXX}
\hline\hline
& & & & & & \\ [-2ex]
$\Sigma^{3}_{\textbf{2}}$ & $\sigma_{2}$ & $\tau_{2}$ & $\sigma_{3}$ & $\tau_{3}$ & $\sigma_{4}$ & $\sigma_{5}$\\ [0.5ex]
\hline
& & & & & & \\ [-2ex]
$|{\rm{Sep}}\rangle$ & $|{\rm{GHZ}}_3^{(1)}\rangle$ & $|{\rm{W}}_3\rangle$ & $|{\rm{GHZ}}_3^{(2)}\rangle$ & $|(333)'_{3}\rangle$ & $|(333)_{4}\rangle$ & $|(333)_{5}\rangle$ \\ [0.5ex]
& $|{\rm{B}}_{i}^{(1)}\rangle_{i=1}^{3}$ & & $|(332)\rangle$ & $|(332)'\rangle$ & &  \\ [0.5ex]
& & & $|(323)\rangle$ & $|(323)'\rangle$ & & \\ [0.5ex]
& & & $|(233)\rangle$ & $|(233)'\rangle$ & &  \\ [0.5ex]
& & & $|(322)\rangle$ & & &  \\ [0.5ex]
& & & $|(232)\rangle$ & & &  \\ [0.5ex]
& & & $|(223)\rangle$ & & &  \\ [0.5ex]
& & & $|{\rm{B}}_{i}^{(2)}\rangle_{i=1}^{3}$ & & &  \\ [0.5ex]
\hline\hline
\end{tabularx}
\label{table:1}
\end{table*}

\begin{figure}[th]
\center{\includegraphics[width=8.8cm]{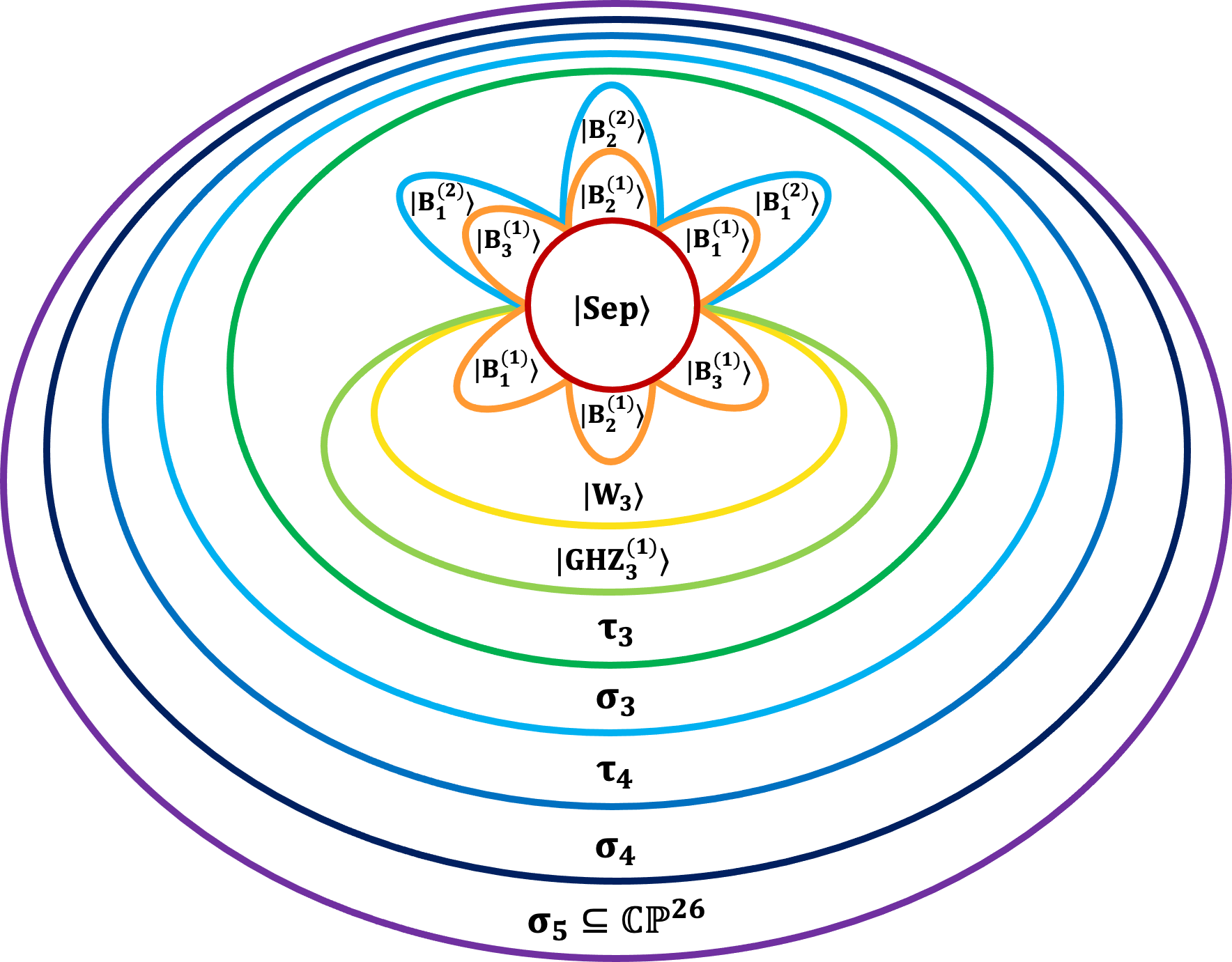}}
\caption{\label{fig:1} (color online). Petal-like classification of SLOCC orbits of three-qutrit states. By noninvertible SLOCC one can go from the outer classes to the inner ones  (from $\sigma_k$ to $\tau_k$ also in an approximate way), thus generating the entanglement hierarchy. Note that states $|{\rm{B}}_i^{(1)}\rangle$ appear with a double petal because to emphasize that they can be obtained starting from either $|{\rm{W}}_3\rangle$ states or $|{\rm{B}}_i^{(2)}\rangle$ states. In contrast, $|{\rm{B}}_i^{(2)}\rangle$ states cannot be obtained from $|{\rm{W}}_3\rangle$ states.}
\end{figure}

From the classification of two-qutrit states (see Appendix \ref{AppxA}), we have biseparable states with one-multirank equal to $(133)$, up to a permutation, as other elements of 
$\sigma_{3}(\Sigma^{3}_{\textbf{2}})$:
\begin{equation}
|{\rm{B}}_{i}^{(2)}\rangle_{i=1}^{3}=\mathcal{P}\{|{\rm{GHZ}}_{2}^{(2)}\rangle|\text{1-qutrit}\rangle\}\,.
\end{equation}
To construct the closure of three-secant variety, i.e., the three-tangent, one can use different limit types at $p_{1}=[1:0:\cdots:0]$.  For instance, we can consider the first limit type which is the addition of Eq. \eqref{W3} with an extra point from the Segre variety (see Ref. \cite{GMO20}). Then, we get
\begin{equation}\label{X3}
|{\rm{X}}_3\rangle=|{\rm{W}}_3\rangle+|\gamma\gamma\gamma\rangle\,,
\end{equation}
where $\alpha\neq\beta\neq\gamma\in\{0,1,2\}$ as an element of $\tau_{3}(\Sigma^{3}_{\textbf{2}})$ with one-multirank equal to $(333)$. 
Indeed, based on the inclusion $\tau_3\subset\sigma_3$, we can conclude that $|{\rm{X}}_3\rangle$ can asymptotically be produced by $|{\rm{GHZ}}_{3}^{(2)}\rangle$. This can be shown by considering the following points:
\begin{equation}
p(\varepsilon)=\frac{1}{\varepsilon}\left((|0\rangle+\varepsilon|1\rangle+\varepsilon|2\rangle)^{\otimes{3}}+\varepsilon|222\rangle-|000\rangle\right)\,.
\end{equation}
For all $\varepsilon\neq{0}$ they correspond to ${\rm{GHZ}}^{(2)}$-type states and indicate a smooth curve in $\sigma_{3}(\Sigma^{3}_{\textbf{2}})$. When $\varepsilon\to{0}$ we have
\begin{equation}
\lim_{\varepsilon\to{0}}p(\varepsilon)=|00\upsilon \rangle+|0\upsilon0\rangle+|\upsilon00\rangle+|222\rangle\,,
\end{equation}
that is equivalent to the state in Eq. \eqref{X3}.

In a similar way, we can also derive as limit process the following states from Eq. (\ref{(332)}), in order to get other elements of $\tau_{3}(\Sigma^{3}_{\textbf{2}})$ with one-multiranks equal to a permutation of $(233)$
\begin{equation}
|{\rm{W}}_3\rangle+\mathcal{P}\{|\alpha\gamma\gamma\rangle\}\,,
\end{equation}
where $\alpha\neq\beta\neq\gamma\in\{0,1,2\}$. Additionally, the states
\begin{equation}
|\alpha\rangle_i|{\rm{GHZ}}_{2}^{(2)}\rangle_{jk}+|\beta\rangle_i|\text{1-qutrit}\rangle^{\otimes2}_{jk}\,,
\end{equation}
where $\{i,j,k\}=\{1,2,3\}$, $\langle{\rm{GHZ}}_{2}^{(2)}|\text{1-qutrit}\rangle^{\otimes2}=0$, and $\langle\alpha|\beta\rangle=0$, have tensor rank and border rank equal to three and four, respectively. So they can be as well considered as elements of $\tau_{3}(\Sigma^{3}_{\textbf{2}})$ with one-multiranks equal to a permutation of $(233)$.

Note that in the three-tangent family, we do not have any element with one-multirank equals to $(223)$ and its permutations. In fact, if one-multirank of a given sate is equal to $(223)$, then the state lives in a smaller tensor product space, here is $\mathbbm{C}^2\otimes\mathbbm{C}^2\otimes\mathbbm{C}^3$, and its border rank is bounded by three \cite{footnote1}, and it is a balanced case \cite{AOP09,BL13}.

The proper four-secant, i.e., the set $\sigma_{4}(\Sigma^{3}_{\textbf{2}})\backslash\sigma_{3}(\Sigma^{3}_{\textbf{2}})$, is the union of the secant hyperplanes $\mathcal{S}_{4}$ represented by Eq. (\ref{S-plane}). For instance, the following states which explicitly come from joining of four distinct points in the Segre variety are elements of $\sigma_{4}(\Sigma^{3}_{\textbf{2}})$ with one-multirank equal to $(333)$
\begin{align}\nonumber
& |000\rangle+|011\rangle+|122\rangle+|221\rangle\,, \\
& |000\rangle+|111\rangle+|122\rangle+|221\rangle\,,
\end{align}
which can be considered as adding two different types of biseparable states $|{\rm{B}}^{(1)}\rangle$, or adding two different types of $|{\rm{GHZ}}_3^{(1)}\rangle$ states, or adding a biseparable state $|{\rm{B}}^{(1)}\rangle$ and a $|{\rm{GHZ}}_3^{(1)}\rangle$ state. Other examples of the proper four-secant family with one-multirank equals to $(333)$ can be considered as joining a general point to the state in Eq. (\ref{GHZ3-2}) as follows:
\begin{equation}\label{GHZ-G}
|{\rm{GHZ}}_{3}^{(2)}\rangle+\mathcal{P}\{|012\rangle\}\,,
\end{equation}
and
\begin{equation}\label{G3}
|{\rm{G}}_3\rangle=|{\rm{GHZ}}_{3}^{(2)}\rangle+|\omega_1\omega_1\omega_1\rangle\,,
\end{equation}
where $|\omega_1\rangle=|0\rangle+|1\rangle+|2\rangle$.

Using Eq. (\ref{Dicke-3-qudit}) one can see that the higher symmetric entangled state
\begin{equation}\label{D(111)}
|{\rm{D}}_{3}^{(1,1,1)}\rangle=|012\rangle+|021\rangle+|102\rangle+|120\rangle+|201\rangle+|210\rangle\,,
\end{equation}
is also an element of $\sigma_{4}(\Sigma^{3}_{\textbf{2}})$ with one-multirank equal to $(333)$. This is because we can relate the above-mentioned symmetric state to the monomial $xyz$ (actually all symmetric states can be related to some homogeneous polynomials since the variables in polynomials are invariant under permutation and each variable can be related to a basis) and we can decompose this monomial as follows
\begin{align}\nonumber
xyz=\frac{1}{24} \Big(  & (x+y+z)^3+(-x-y+z)^3
\\ \label{xyz}
& +(-x+y-z)^3+(x-y-z)^3\Big)\,.
\end{align}
So, using the Dirac notation, we can rewrite the state $|{\rm{D}}_{3}^{(1,1,1)}\rangle$ in Eq. (\ref{D(111)}) based on the above decomposition as follows:
\begin{equation}
|{\rm{D}}_{3}^{(1,1,1)}\rangle=\frac{1}{4}\Big(|\omega_1\rangle^{\otimes 3}+|\omega_2\rangle^{\otimes 3}+|\omega_3\rangle^{\otimes 3}+|\omega_4\rangle^{\otimes 3}\Big)\,,
\end{equation}
where $|\omega_2\rangle=-|0\rangle-|1\rangle+|2\rangle$, $|\omega_3\rangle=-|0\rangle+|1\rangle-|2\rangle$, and $|\omega_4\rangle=|0\rangle-|1\rangle-|2\rangle$. Hence, both the tensor rank and the border rank of this state are four.

In the four-secant family, we do not have any element with one-multirank equals to $(233)$ and its permutations. Indeed, if one-multirank of a given sate is equal to $(233)$ then the state lives in $\mathbbm{C}^2\otimes\mathbbm{C}^3\otimes\mathbbm{C}^3$, and its border rank is bounded by three, but tensor rank can be three or four \cite{BL13}.

Concerning the closure of the four-secant variety, i.e., the four-tangent, we are led to consider $\sigma_{4}(\Sigma^{3}_{\textbf{2}})\setminus\sigma_{3}(\Sigma^{3}_{\textbf{2}})$ locally closed. In fact, all $(333)$ trilinear forms in Ref. \cite{Allums} have been proved numerically \cite{BS20} and analytically \cite{Bernardi} to have both tensor rank and border rank equal to four \cite{footnote2}.

Although any general state of three-qutrit system that has a nonzero determinant of matrix $\mathcal{F}$ in Eq. (\ref{strassen}) can be considered as an element of proper five-secant family, the following state which explicitly comes from joining of five distinct points in the Segre variety and obeys Eq. (\ref{S-plane}), is an element of $\sigma_{5}(\Sigma^{3}_{\textbf{2}})$ with one-multirak equal to $(333)$
\begin{equation}
|{\rm{G}}_{3}\rangle+t~(|1\rangle+|2\rangle)\otimes(|0\rangle+|2\rangle)\otimes(|0\rangle+|1\rangle)\,,
\end{equation}
where $t\in\mathbbm{C}\setminus\{0,1\}$. The determinant of the matrix $\mathcal{F}$ for this state is $2t(1-t)$. Note that for $t=1$ the border rank is four and the tensor rank is also four, so the state belongs to the four-secant family in this case.

Since the highest tensor rank for a three-qutrit state is five \cite{BH13}, we do not need to construct the Zariski closure of the five-secant family.

It is worth noting that in the classification of three-qutrit states, all the elements in four- and five-secant families are genuinely entangled.

To have an exhaustive classification, we have written each subfamily of three-, four-, and five-secant families in terms of their one-multiranks in Table \ref{table:1}. Also, we have used a prime for the states in tangent to discriminate secant and tangent families where they have same one-multiranks. In addition, we have put a subscript $k$ to indicate members appearing in different $k$-secant families with the same one-multirank.

In summary, this classification provides us five secant families (seven secant/tangent families), and 22 subfamilies (Table \ref{table:1}). These classes are pictorially represented in Fig.\ref{fig:1}. {Obviously, a finer classification can be obtained by utilizing an extra SLOCC invariant (see Appendix \ref{AppxB}).}

\section{Generalizations}\label{Sec.iv}

We generalize here some of the results found in the previous section to tripartite $\mathbbm{C}^d\otimes\mathbbm{C}^d\otimes\mathbbm{C}^d$ systems as well to $n$-qudit systems.

As one can see, going beyond qubit, there are several types of $\rm{GHZ}$ states (see for instance, Eqs. (\ref{GHZ3-1}) and (\ref{GHZ3-2})). This is because we have different types of excitations rather than qubit systems. So we can draw the following conclusions for $d,n\geq{3}$:
\begin{equation}\label{GHZn-z}
|{\rm{GHZ}}_{n}^{(\zeta)}\rangle=|\alpha_1\rangle^{\otimes{n}}+\cdots+|\alpha_{\zeta+1}\rangle^{\otimes{n}}\in\sigma_{\zeta+1}(\Sigma^{n}_{\textbf{d-1}})\,,
\end{equation}
where $1\leq\zeta\leq{d-1}$ and $\alpha_i\neq\alpha_j\in\{0,1,\ldots,d-1\}$. Then, based on Eq. (\ref{GHZn-z}), we can create $(n-m+1)$-separable states as follows:
\begin{equation}
\mathcal{P}\{|{\rm{GHZ}}_{m}^{(\zeta)}\rangle|\text{1-qudit}\rangle^{\otimes{n-m}}\}\in\sigma_{\zeta+1}(\Sigma^{n}_{\textbf{d-1}})\,,
\end{equation}
where $2\leq{m}\leq{n-1}$.

From Eq. (\ref{W3}), we can draw the following conclusion for $d,n\geq{3}$:
\begin{align}\nonumber
|{\rm{W}}_n\rangle&=|{\rm{D}}_{n}^{\mathcal{P}(n-1,1,0,\cdots,0)}\rangle=\sum_{i}\mathcal{P}_{i}\{|\alpha\rangle^{\otimes{n-1}}\otimes|\beta\rangle\} \\ \label{Wn}
&\in\tau_{2}(\Sigma^{n}_{\textbf{d-1}})\,,
\end{align}
where $\alpha\neq\beta\in\{0,1,\ldots,d-1\}$ and
\begin{equation}\label{Dicke-n-qudit}
|{\rm{D}}_{n}^{\jmath}\rangle=\sqrt{\frac{\prod_{i}{j_i!}}{n!}}\sum_{\pi\in\mathfrak{S}_{n}}\pi\{|0\rangle^{\otimes{j_1}}\otimes\cdots\otimes|d-1\rangle^{\otimes{j_d}}\}\,,
\end{equation}
are the so-called $n$-qudit Dicke states, with excitations shown as $\jmath=(j_1,\ldots,j_d)$ where $j_1+\cdots+j_d=n$.

Furthermore, from Eq. (\ref{X3}) we can conclude, for $d,n\geq{3}$:
\begin{equation}\label{Xn}
|{\rm{X}}_n\rangle=|{\rm{W}}_n\rangle+|\gamma\gamma\gamma\rangle\in\tau_{3}(\Sigma^{n}_{\textbf{d-1}})\,,
\end{equation}
where $\gamma\in\{0,1,\ldots,d-1\}$ is different from $\alpha$ and $\beta$ in Eq. (\ref{Wn}).

For $d$-qudit states we have the following results, which respectively comes from Eqs. (\ref{GHZ-G}) and (\ref{G3}),
\begin{align}
&|{\rm{GHZ}}_{d}^{(d-1)}\rangle+\mathcal{P}\{|01\cdots(d-1)\rangle\}&\in\sigma_{d+1}(\Sigma^{d}_{\textbf{d-1}})\,, \\
&|{\rm{G}}_{d}\rangle=|{\rm{GHZ}}_{d}^{(d-1)}\rangle+|\Omega\rangle^{\otimes{d}}&\in\sigma_{d+1}(\Sigma^{d}_{\textbf{d-1}})\,,
\end{align}
where $|\Omega\rangle=|0\rangle+\cdots+|d-1\rangle$.

Let us now indulge on Dicke states. Since they are symmetric, i.e.,  are invariant under any permutation of the parties, they correspond to monomials, up to scaling the variables. Thus, their tensor rank can be computed as the Waring rank of the corresponding monomials. The Waring rank of a homogeneous $d$-variate degree-$n$ polynomial $f\in\mathbbm{C}[x_1,\ldots,x_d]_n$ is the minimum number of terms contained in $f$ when it is expressed as a combination of $n^{\text{th}}$ powers of linear forms. The Waring rank of a general monomial has been solved in Ref. \cite{CCG12}.

\begin{manualtheorem}{1}{(Ref. \cite{CCG12})}\label{theo1}
The Waring rank of a monomial $x_0^{d_0}\cdots x_n^{d_n}$ with $0<d_0\leq\cdots\leq d_n$ is equal to $\prod_{i=1}^{n}(d_i+1)$.
\end{manualtheorem}

In addition, the conjecture in Ref. \cite{Oeding20} provides the border rank of a general monomial.

\begin{manualconjecture}{1}{(Ref. \cite{Oeding20})}\label{conj1}
The border rank of a monomial $x_0^{d_0}\cdots x_n^{d_n}$ with $0<d_0\leq\cdots\leq d_n$ is equal to $\prod_{i=0}^{n-1}(d_i+1)$.
\end{manualconjecture}

From Theorem \ref{theo1} and Conjecture \ref{conj1} we can derive the further general results about qutrit Dicke states (and in passing also for multiqubit Dicke states, see Appendix \ref{AppxC}).

\begin{manualproposition}{1}
For $d\geq{3}$, there is no symmetric entangled state in the higher secant variety of $\mathbbm{P}^{d^3-1}$.
\end{manualproposition}
{\it Proof.}
The generic symmetric rank of a tensor in ${\rm{Sym}}^n\mathbbm{C}^d$ is equal to the expected symmetric tensor rank which is
\begin{equation}\label{SymGenericRank}
\left\lceil\frac{\binom{n+d-1}{n}}{d}\right\rceil\,,
\end{equation}
except for (i) $n=2$ where it is equal to $d$, and (ii) the pairs $(n,d)=(3,5),(4,3),(4,4),(4,5)$ where the generic symmetric rank is equal to the expected plus one \cite{AH95, BO08}. Comparing Eq. (\ref{SymGenericRank}) with $n=3$ to Eq. (\ref{TripartiteGenericRank}) provides the proof of the proposition.
\qed

\begin{manualproposition}{2}
For $n\geq{4}$ qutrits, Dicke states are not in the higher secant variety of $\mathbbm{P}({\rm{Sym}}^n\mathbbm{C}^3)$.
\end{manualproposition}
{\it Proof.}
Based on Theorem \ref{theo1} and Conjecture \ref{conj1}, for an $n$-qutrit Dicke state, the maximum border rank achieved when $\jmath=(\lceil\frac{n}{3}\rceil,\lfloor\frac{n}{3}\rfloor,n-\lceil\frac{n}{3}\rceil-\lfloor\frac{n}{3}\rfloor)$ in Eq. (\ref{Dicke-n-qudit}). So,
\begin{align}\nonumber
&|{\rm{D}}_{n}^{(\lceil\frac{n}{3}\rceil,\lfloor\frac{n}{3}\rfloor,n-\lceil\frac{n}{3}\rceil-\lfloor\frac{n}{3}\rfloor)}\rangle\in\\
&\begin{cases}
  \sigma_{(\lfloor\frac{n}{3}\rfloor+1)(n-\lceil\frac{n}{3}\rceil-\lfloor\frac{n}{3}\rfloor+1)}(\Sigma^{n}_{\textbf{2}}) & \text{if}~n=3i~(i\in\mathbbm{N})\,, \\
  \tau_{(\lfloor\frac{n}{3}\rfloor+1)(n-\lceil\frac{n}{3}\rceil-\lfloor\frac{n}{3}\rfloor+1)}(\Sigma^{n}_{\textbf{2}})  & \text{otherwise}\,.
\end{cases}
\end{align}
On the other hand, the generic symmetric rank of a tensor in ${\rm{Sym}}^n\mathbbm{C}^3$ is equal to $\left\lceil\frac{(n+1)(n+2)}{6}\right\rceil\,,$ except for $n=4$ where it is six. Hence, in contrast to multiqubit Dicke states, multiqutrit Dicke states are not in the higher secant variety in $\mathbbm{P}({\rm{Sym}}^n\mathbbm{C}^3)$.
\qed

Moreover, since there is no symmetric entangled state in the higher secant family of $3$-qutrit systems, it turns out that for $n\geq{3}$ qutrits, there is no symmetric entangled state in the higher secant variety. 

For general multipartite systems in $\mathbbm{C}^{d_1}\otimes\cdots\otimes\mathbbm{C}^{d_n}$, there is a conjecture in Ref. \cite{AOP09}, claiming that the generic tensor rank is equal to the expected tensor rank that is
\begin{equation}
\left\lceil\frac{\prod_{i=1}^n d_i}{\sum_{i=1}^n d_i-n+1}\right\rceil\,,
\end{equation}
except for $\mathbbm{C}^{4\times{4}\times{3}}$, $\mathbbm{C}^{(2i+1)\times(2i+1)\times{3}}$, and $\mathbbm{C}^{(i+2)\times(i+2)\times{2}\times{2}}$, with $i=1,2,\ldots$. In this exceptional cases the generic tensor rank is equal to the expected plus one. So, based on this conjecture, it is also possible to classify entanglement in multipartite systems.

\section{Conclusion and Outlook}\label{Sec.v}

Using algebraic-geometric tools, we studied entanglement characterization of three-qudit $\mathbbm{C}^d\otimes\mathbbm{C}^d\otimes\mathbbm{C}^d$ systems. Specifically, we used secant varieties and one-multiranks that are SLOCC invariants, to present entanglement classification of three-qudit entanglement as a generalization to our previous work in Ref. \cite{GMO20}. As a prominent example we have provided a fine-structure classification for three-qutrit pure states. This  can be considered as the core  classification of tripartite $\mathbbm{C}^d\otimes\mathbbm{C}^d\otimes\mathbbm{C}^d$ states as well as ($n\geq{4}$)-qutrit states. Indeed, with this method, one can always use $n$-qudit classification as a partial classification of $(n + 1)$-qudit systems. Outside this core classification, the results for larger systems ($d>3$ and/or $n>3$) have been derived by also relying on conjectures of tensor theory.

Not only is our classification operationally meaningful as it quantifies entanglement as a resource but also this classification can be seen in terms of the order of entanglement strength from Segre variety that contains no entanglement, to the higher secant family. Indeed, the tools we have used for entanglement characterization, i.e., tensor rank and border rank, can be seen as the generalized Schmidt rank and its counterpart. More precisely, the Schmidt measure that quantify entanglement of a multipartite state $|\psi\rangle$ can be defined as the logarithm of the rank of the tensor $\psi$. On the other hand, generic tensor rank can be considered as a discrete measure of entanglement. Based on this fact, one can conclude that symmetric states are much less entangled than general states. Although we have shown this fact for multiqubit systems in Ref. \cite{GMO20}, and for three-qudit and multiqutrit systems in Sec. \ref{Sec.iv},  this is a general fact since generic symmetric tensor rank has a polynomial growth while generic tensor rank has an exponential growth.

Along the potential applications mentioned in Ref. \cite{GMO20} that can also be considered for the higher dimensional systems, it is captivating that this kind of classification can also be considered as a reference to study SLOCC and asymptotic SLOCC interconversions among different resources based on tensor rank \cite{CDS08, YCGD10,CCDJW10} and border rank \cite{YGD14,VC15}, respectively.

It would be also desirable to extend the proposed classification method to mixed states. This goal will be pursued starting from possible connections with the Schmidt number vector classification of Refs. \cite{HD13, HPD13}. Indeed, the idea is utilizing the generalization of Schmidt rank for pure states to Schmidt number for mixed states \cite{TH00}. So, the Schmidt rank vector is nothing but the multirank we have used in our method and the Schmidt number vector is a tuple of digits obtained from a particular ensemble decomposition of a given mixed state \cite{HD13}.

\section*{Acknowledgments}
The authors warmly thank Giorgio~Ottaviani for many insightful discussions and invaluable comments. They would also like to thank Fr\'ed\'eric~Holweck for a preliminary reading of the ms. M.~G.  acknowledges fruitful discussions with Alessandra~Bernardi, Jaros\l{}aw~Buczy\'{n}ski, Pedram~Karimi, Luke~Oeding and Adam~Sawicki.

\appendix

\section{Fine-structure classification of two-qutrit entanglement}\label{AppxA}
In this Appendix, we provide a full entanglement classification for two-qutrit states which can be used as the core for the entanglement classification of three-qutrit states.

For two-qutrit states, the Segre twofold $\Sigma^{2}_{\textbf{2}}\subset\mathbbm{P}^{8}$, i.e., the set of fully separable states of two qutrits, consists of general points given by affine coordinates $p=[1:a:b:c:ac:bc:d:ad:bd]$ where $a$, $b$, $c$, and $d$ are complex parameters and one or more parameters can tend to infinity.

Moving on to the proper two-secant variety, i.e., the union of the secant planes $\mathcal{S}_{2}=\lambda_1p_1+\lambda_2p_2$, we have generic elements given by the following coordinates:
\begin{align}\nonumber
[& \lambda_1+\lambda_2:\lambda_1a_1+\lambda_2a_2:\lambda_1b_1+\lambda_2b_2:\lambda_1c_1+\lambda_2c_2: \\ \nonumber
& \lambda_1a_1c_1+\lambda_2a_2c_2:\lambda_1b_1c_1+\lambda_2b_2c_2:\lambda_1d_1+\lambda_2d_2: \\
& \lambda_1a_1d_1+\lambda_2a_2d_2:\lambda_1b_1d_1+\lambda_2b_2d_2]\,.
\end{align}
It is easy to see that $[1:0:0:0:1:0:0:0:0]$ is an elements of $\sigma_2(\Sigma^{2}_{\textbf{2}})$. Actually, the following general state:
\begin{equation}\label{GHZ2-1}
|{\rm{GHZ}}_{2}^{(1)}\rangle=|\alpha\alpha\rangle+|\beta\beta\rangle\,,
\end{equation}
where $\alpha\neq\beta\in\{0,1,2\}$, can represent all elements of proper two-secant family with one-multiranks equal to $(22)$.

Obviously, one can rewrite the secant planes as $\mathcal{S}_{2}=p_1+\mu(p_2-p_1)$ where 
$\lambda_1=1-\mu$ and $\lambda_2=\mu$. Now, we consider the situation where second point tends to the first one, i.e., $p_2\to{p_1}$, by taking $p_2(\epsilon)=[1:a_1+\epsilon:b_1+\epsilon:c_1+\epsilon:(a_1+\epsilon)(c_1+\epsilon):(b_1+\epsilon)(c_1+\epsilon):d_1+\epsilon:(a_1+\epsilon)(d_1+\epsilon):(b_1+\epsilon)(d_1+\epsilon)]$. This will give us the coordinates of the elements in the two-tangent variety which can be considered as the closure of two-secant variety. However, for two-qutrit states, we have the special situation that all points on the tangent, i.e.,
\begin{align}\nonumber
p'=&\lim_{\epsilon\rightarrow 0}\left(p_1+\frac{\mu}{\epsilon}\big(p_2(\epsilon)-p_1\big)\right)=[1:a_1+\mu:b_1+\mu: \\ \nonumber
& c_1+\mu:a_1c_1+\mu(a_1+c_1):b_1c_1+\mu(b_1+c_1): \\
& d_1+\mu:a_1d_1+\mu(a_1+d_1):b_1d_1+\mu(b_1+d_1)]\,,
\end{align}
lie also on the proper two-secant since
\begin{align}\nonumber
p'=[& 1:a_1+\mu:b_1+\mu:c_1+\mu:(a_1+\mu)(c_1+\mu): \\ \nonumber
& (b_1+\mu)(c_1+\mu):d_1+\mu:(a_1+\mu)(d_1+\mu): \\
& (b_1+\mu)(d_1+\mu)]-\mu^2[0:0:0:0:1:1:0:1:1]\,,
\end{align}
which explicitly comes from joining of two distinct point in the Segre variety, i.e., superposition of two fully separable states. It means that the proper two-secant is closed.

The proper three-secant, i.e., the set $\sigma_3(\Sigma^{2}_{\textbf{2}})/\sigma_2(\Sigma^{2}_{\textbf{2}})$,  is the union of the secant hyperplanes $\mathcal{S}_3$ represented by Eq. (\ref{S-plane}). Indeed, joining of three distinct points in the Segre variety gives rise to elements of three-secant family. For instance,
\begin{equation}\label{GHZ2-2}
|{\rm{GHZ}}_{2}^{(2)}\rangle=|00\rangle+|11\rangle+|22\rangle\,,
\end{equation}
is an element of $\sigma_3(\Sigma^{2}_{\textbf{2}})$ with one-multirank equals to $(33)$. In a similar way to two-secant variety, one can see that the proper three-secant is locally closed.

\begin{table}[t]
\centering
\caption{Fine-structure classification of two-qutrit entanglement.}
\vspace{8pt}
\begin{tabularx}{\linewidth}{XXX}
\hline\hline
& & \\ [-2ex]
$\Sigma^{2}_{\textbf{2}}$ & $\sigma_{2}$ & $\sigma_{3}$ \\ [0.5ex]
\hline
& & \\ [-2ex]
$|\rm{Sep}\rangle$ & $|{\rm{GHZ}}_{2}^{(1)}\rangle$ & $|{\rm{GHZ}}_{2}^{(2)}\rangle$ \\ [0.5ex]
\hline\hline
\end{tabularx}
\label{table:2}
\end{table}

Briefly, this classification provide us three secant families that coincide with the three SLOCC classes, namely, separable and two inequivalently entangled states that come from superposition of two and three fully separable states (Table \ref{table:2}).

Already from this classification we can draw a general conclusion. That is, for $n\geq{2}$ qutrits we have
\begin{align}\label{general-qutrit}
&\mathcal{P}\{|{\rm{GHZ}}_{2}^{(1)}\rangle|\text{1-qutrit}\rangle^{\otimes{(n-2)}}\} \in \sigma_2(\Sigma^{n}_{\textbf{2}})\,, \\
&\mathcal{P}\{|{\rm{GHZ}}_{2}^{(2)}\rangle|\text{1-qutrit}\rangle^{\otimes{(n-2)}}\} \in \sigma_3(\Sigma^{n}_{\textbf{2}})\,,
\end{align}
where $\mathcal{P}\{\cdot\}$ denotes all possible permutations of subsystems.

\section{Finer classification of three-qutrit entanglement}\label{AppxB}

Since the Schmidt measure can be defined as the logarithm of the tensor rank of a quantum state, one can conclude that tensor rank is itself an SLOCC invariant. Therefore, we can employ it to improve the classification algorithm by eventually splitting subfamilies into sub-subfamilies with the same tensor rank. Although determining the tensor rank of a given quantum state is NP hard \cite{Haastad90}, it 
could also results a useful tool for studying SLOCC interconversions among specific quantum states.

In Ref. \cite{BHO14}, a classification of three-qutrit entanglement is presented in five families according to the description of fundamental invariants provided in Refs. \cite{Nurmiev1,Nurmiev2}. It is also determined which fundamental invariants of ${\rm{SL}}(3,\mathbbm{C})^{\times{3}}$ vanish on tensors for each possible tensor rank. Here, we utilize tensor rank as an extra SLOCC invariant to present a finer classification of  three-qutrit entanglement with respect to the classification presented in Table \ref{table:1}, such that it contains the information of Ref. \cite{BHO14}.

\begin{figure}[t]
\center{\includegraphics[width=5cm]{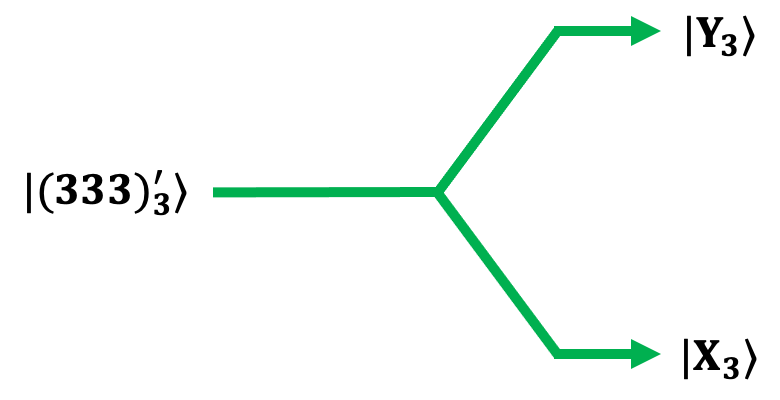}}
\caption{\label{fig:2} Pictorial representation of the fact that using tensor rank as the third SLOCC invariant, the subfamily $|(333)'_3\rangle$ of Table \ref{table:1} can be split into two sub-subfamilies $|{\rm{X}}_3\rangle$ and $|{\rm{Y}}_3\rangle$ with tensor ranks equal to four and five, respectively.}
\end{figure}

To this end, consider the following state:
\begin{equation}\label{Y3}
|{\rm{Y}}_3\rangle=|002\rangle+|020\rangle+|200\rangle+|011 \rangle+|101\rangle+|110\rangle\,,
\end{equation}
and the following points:
\begin{equation}
q(\varepsilon)=\frac{(|0\rangle+\varepsilon|1\rangle+\varepsilon^{2}|2\rangle)^{\otimes{3}}+(|0\rangle-\varepsilon|1\rangle)^{\otimes{3}}-2|000\rangle}{\varepsilon^2}\,,
\end{equation}
that for all $\varepsilon\neq{0}$ correspond to ${\rm{GHZ}}^{(2)}$-type states. When $\varepsilon\to{0}$ we have
\begin{equation}\label{limY3}
\lim_{\varepsilon\to{0}}q(\varepsilon)=|002\rangle+|020\rangle+|200\rangle+2(|011 \rangle+|101\rangle+|110\rangle)\,,
\end{equation}
that is equivalent to the state in Eq. \eqref{Y3}. So $|{\rm{Y}}_3\rangle$ can be considered as another element of $\tau_{3}(\Sigma^{3}_{\textbf{2}})$ with one-multirank equal to $(333)$. Moreover, it can asymptotically be obtained from $|{\rm{GHZ}}_{3}^{(2)}\rangle$. It worth noting that the states in Eqs. \eqref{X3} and \eqref{Y3} are not equivalent since the tensor rank of the former is four, while of the later is five. In fact, we can rewrite Eq. \eqref{Y3} as follows:
\begin{align}\nonumber
&|{\rm{Y}}_3\rangle=\frac{1}{3}\left[(2|0\rangle+|2\rangle)^{\otimes{3}}-2(|0\rangle+|2\rangle)^{\otimes{3}}+|222\rangle\right] \\
&+\frac{1}{2\sqrt{3}i}\left[(2\xi+1)|0\rangle-|1\rangle)^{\otimes{3}}-((2\xi^2+1)|0\rangle-|1\rangle)^{\otimes{3}}\right],
\end{align}
with $\xi=\exp(2\pi i / 3)$. Hence, using the tensor rank as the third SLOCC invariant, we can split the subfamily $|(333)'_3\rangle\in\tau_{3}(\Sigma^{3}_{\textbf{2}})$ into two sub-subfamilies with tensor ranks equal to four and five, respectively (see Fig. \ref{fig:2}).

\section{More on the classification of $n$-qubit Dicke states}\label{AppxC}

Regarding Theorem \ref{theo1} and Conjecture \ref{conj1}, we have the following result for the $n$-qubit Dicke states $|{\rm{D}}_{n}^l\rangle$ (with $l$ excitations). If $1\leq{l}<\lfloor\frac{n}{2}\rfloor$, tensor rank and border rank are equal to $n-l+1$ and $l+1$, respectively. For $l=\lfloor\frac{n}{2}\rfloor$, we have two situations; (1) if $n=\text{even}$, tensor rank and border rank are both equal to $\frac{n}{2}+1$, and (2) if $n=\text{odd}$, tensor rank and border rank are equal to $\lceil\frac{n}{2}\rceil+1$ and $\lfloor\frac{n}{2}\rfloor+1$, respectively. Hence, the relation between tensor rank and border rank of $n$-qubit Dicke states is as follows:
\begin{equation}
\text{Tensor rank}\left(|{\rm{D}}_{n}^l\rangle\right)+\text{Border rank}\left(|{\rm{D}}_{n}^l\rangle\right)=n+2\,.
\end{equation}
Based on this fact, we draw the following result
\begin{equation}
|{\rm{D}}_{n}^{\lfloor\frac{n}{2}\rfloor}\rangle\in
\begin{cases}
  \sigma_{\frac{n}{2}+1}(\Sigma^{n}_{\textbf{1}}) & \text{if $n=$ even}\,, \\
  \tau_{\lfloor\frac{n}{2}\rfloor+1}(\Sigma^{n}_{\textbf{1}})  & \text{if $n=$ odd}\,.
\end{cases}
\end{equation}
Therefore, for an even number of qubits, regarding the rank and border rank information the Dicke state $|{\rm{D}}_{n}^{\frac{n}{2}}\rangle$ is in the proper $(\frac{n}{2}+1)$-secant family while based on the higher derivative information it is in the osculating hyperplane that we take it in the tangent family. Geometrically, it means that this special state is in the intersection of proper $(\frac{n}{2}+1)$-secant family and $(\frac{n}{2}+1)$-tangent family.




\begin{thebibliography}{99}
\bibitem{example} {For example, for three qutrits where the partition is 1|23,
let us consider two states, one from the family span$\{\Psi_1,\Psi_2\}$ and another from the family span$\{\Psi_0,\Psi_0,\Psi_2\}$ as follow:
\begin{align*}
&|000\rangle+|011\rangle+|022\rangle+|101\rangle+|112\rangle= \\
&|0\rangle(|00\rangle+|11\rangle+|22\rangle)+|1\rangle(|01\rangle+|12\rangle)\,,
\end{align*}
and
\begin{align*}
&|000\rangle+|011\rangle+|022\rangle+|101\rangle+|202\rangle= \\
&|0\rangle(|00\rangle+|11\rangle+|22\rangle)+|1\rangle|01\rangle+|2\rangle|02\rangle\,.
\end{align*}
Now, let us consider the partition 12|3 for the above states. We will have
\begin{align*}
&|000\rangle+|011\rangle+|022\rangle+|101\rangle+|202\rangle= \\
&|00\rangle|0\rangle+(|01\rangle+|10\rangle)|1\rangle+(|02\rangle+|11\rangle)|2\rangle\,,
\end{align*}
and
\begin{align*}
&|000\rangle+|011\rangle+|022\rangle+|101\rangle+|202\rangle= \\
&|00\rangle|0\rangle+(|01\rangle+|10\rangle)|1\rangle+(|02\rangle+|20\rangle)|2\rangle\,.
\end{align*}
As we see by changing the partition and algorithm two states from different families go to the family span$\{\Psi_0,\Psi_1,\Psi_1\}$ where $\Psi_0$ denotes a fully separable state while $\Psi_1$ and $\Psi_2$ denote two inequivalent fully inseparable states. Therefore, there is an overlap between some families of three-qutrit entanglement, and thus this approach cannot be
used to exactly identify to which family a given three-qutrit
state belongs to.}
\bibitem{footnote1}
{Let us consider two cases which have one-multiranks equal to a permutations of $(223)$: (1) Concerning Eq. (\ref{(322)}), one can consider the states $|{\rm{W}}_3\rangle+|\alpha\beta\gamma\rangle$. It is obvious that we can write these states as $|\alpha\alpha\beta\rangle+|\alpha\beta\rangle(|\alpha\rangle+|\gamma\rangle)+|\beta\alpha\alpha\rangle$ which clearly have tensor rank and border rank equal to three. (2) With a better choice of basis one can also consider the sates $|{\rm{W}}_3\rangle+|\beta\beta\gamma\rangle$. These states can be easily written as 
$|\alpha\alpha\rangle(-|\alpha\rangle+|\beta\rangle)+(|\alpha\rangle+|\beta\rangle)(|\alpha\rangle+|\beta\rangle)|\alpha\rangle+|\beta\beta\rangle(-|\alpha\rangle+|\gamma\rangle)$ which clearly have tensor rank and border rank equal to three.}
\bibitem{footnote2}
{Actually, the list in Ref. \cite{Allums} results incomplete, however the forms not present there are conjectured to have as well tensor rank and border rank equal to four \cite{Bernardi}.}
\end{thebibliography}
\end{document}